\documentclass{sig-alternate-10pt}

\usepackage[pass]{geometry}

\usepackage{epsfig,endnotes}
\usepackage{epstopdf}
\usepackage{times}
\usepackage{color}
\usepackage{subfigure}
\usepackage{url}
\usepackage{graphicx}
\usepackage{soul}
\usepackage[ruled,vlined,linesnumbered]{algorithm2e}
\usepackage{xspace}
\usepackage[usenames,dvipsnames]{xcolor}
\usepackage{flushend}  
\usepackage{etoolbox}

\newcommand{\fig}{{Fig.}}
\newcommand\comment[1]{}
\newcommand{\eg}{{e.g.,}\xspace}

\newcommand{\folder}{./fig}

\newcommand{\hh}{{H2}\xspace}

\newcommand{\h}{{H1}\xspace}
\newcommand{\quic}{{QUIC}\xspace}
\newcommand{\spd}{{SPDY}\xspace}
\newcommand{\spdy}{{SPDY}\xspace}
\newcommand{\urldata}{\cite{our_page}}   
\newcommand{\pl}{{PlanetLab}\xspace}

\begin{document}

\date{}

\title{To HTTP/2, or Not To HTTP/2, That Is The Question}
\numberofauthors{1}
\author{
    \alignauthor 
Matteo Varvello$^\star$, 
Kyle Schomp$^\dagger$, 
David Naylor$^\ddagger$, 
Jeremy Blackburn$^\star$,\\
Alessandro Finamore$^\star$,
Kostantina Papagiannaki$^\star$
\\
\affaddr{$^\star$Telef\'onica Research},
\affaddr{$^\dagger$Case Western Reserve University},
\affaddr{$^\ddagger$Carnegie Mellon University} 
}

\maketitle

\begin{abstract}

As of February, 2015, HTTP/2, the update to the 16-year-old HTTP 1.1, is
officially complete. HTTP/2 aims to improve the Web experience by solving
well-known problems (\eg head of line blocking and redundant headers), while
introducing new features (\eg server push and content priority). On paper
HTTP/2 represents the future of the Web. Yet, it is unclear whether the Web
itself will, and should, hop on board. To shed some light on these questions,
we built a measurement platform that monitors HTTP/2 adoption and performance
across the Alexa top 1 million websites on a daily basis. Our system is live
and up-to-date results can be viewed at \urldata. In this paper, we report our
initial findings from a 6 month measurement campaign (November 2014--May
2015). We find 13,000 websites \emph{reporting} HTTP/2 support, but only 600,
mostly hosted by Google and Twitter, actually \emph{serving} content. In terms
of speed, we find no significant benefits from HTTP/2 under stable network
conditions. More benefits appear in a 3G network where current Web development
practices make HTTP/2 more resilient to losses and delay variation than
previously believed.

\end{abstract}

\section{Introduction}
\label{sec:intro}

HTTP/2 (\hh for short) is the new version of HTTP, expected to replace version
1.1 (H1) which was standardized in 1999. \hh promises to make the Web faster and
more efficient by compressing headers, introducing server push, fixing  head of
line blocking, and loading page elements in parallel over a single TCP
connection (cf.~Section~\ref{sec:background}). Although the IETF does not
require encrypting \hh connections with TLS (as originally proposed), the major
browser vendors have decided to only support encrypted \hh traffic.

While on paper \hh represents the future of the Web, it is unclear whether its
adoption will face a struggle similar to IPv6. As discussed
in~\cite{akhshabi_2011SIGCOMM}, the adoption of a new protocol in presence of
ossification largely depends on the ratio between its benefits and its costs.
Modern websites are designed to deal with \h's inefficiencies, employing hacks
like spriting, inlining, etc.~\cite{h1_acks}.  While \hh would remove the need
for such hacks, in theory simplifying Web development, it is unclear how much
performance improvement it can provide compared to \h coupled with the hacks
above. Furthermore, given the large adoption of such hacks, it is unclear if/how
they can coexist with \hh (vital for the adoption of \hh as one cannot expect
web developers to rebuild their websites overnight).

Motivated by the above, in this work we build a measurement platform which
monitors \hh adoption and assesses its performance. Our measurement platform
consists of a \emph{master} and several \emph{workers}.  The master is a Linux
server that coordinates measurements and collect results. The workers are both
\pl machines and machines within our lab. We use \pl to probe Alexa's top
1~million websites daily looking for claims of \hh support. Then, we use our own
machines to verify whether sites claiming \hh support actually do and, if so, to
analyze which features they use and end-user experience. Results are published
daily at \urldata.

This paper reports findings from a 6 months measurement campaign, from November
2014 until May 2015 (cf.~Section~\ref{sec:results}). Overall, we find that
13,000 websites \emph{report} \hh support, but only about 600, mostly hosted by
Google and Twitter, actually \emph{serve} content over \hh on a consistent
basis. A macroscopic analysis of such content suggests that there was no
significant content modification as websites adopt \hh. A more in-depth analysis
also unveils that classic \h hacks are still present with \hh. For example,
inlining (putting CSS styles and JavaScript code directly in HTML) is still
widely used, reducing caching benefits at the browser; the same is true of
domain sharding (spreading Web objects across several domains either on purpose
or because needed) which causes only a 15\% reduction of TCP connections used
with \hh. Interestingly, domain sharding makes \hh more resilient to losses and
delay variation typical of mobile networks.

\section{Background and Related Work}
\label{sec:background}

\noindent{\bf \h} is an \emph{ASCII} protocol that allows a client to
request/submit content from/to a server.  \h is mostly used to fetch web pages,
where clients request \emph{objects} from a server and the resulting response is
serialized over a persistent TCP connection. Pipelining requests improves page
load time for pages with many objects, but the benefits are limited since \h
requires servers to serve requests in order. Thus, an early request for a large
object can delay all subsequent pipelined requests (head of line blocking).
Clients mitigate this by opening several concurrent TCP connections to a server,
which incurs additional overhead (TCP state on the server, TCP handshake
latency, and TLS session setup in the case of HTTPS~\cite{naylor_conext_2014}).
Accordingly, browser vendors set a limit on the maximum number of open
connections to a domain, \eg 6 in Chrome and 15 in
Firefox~\cite{chrome_connections}. Web developers have dealt with this
limitation through the use of \emph{domain sharding}, where content is
distributed across different domains, circumventing the single domain connection
limit. Finally, \h requires the explicit transmission of headers on a per
request/response basis. Therefore, common headers (like server version) are
retransmitted with each object, causing high overhead especially for pages with
a large number of small embedded objects.\\

\noindent{\bf \spd} is a Google protocol for delivering web content.  It is a
fully binary protocol, enabling efficient parsing, lighter network footprint,
and, most importantly, reducing exposure to security issues resulting from the
failure to properly sanitize input values. \spd opens a single TCP connection to
a domain and \emph{multiplexes} requests and responses, called \emph{streams},
over that connection, which reduces TLS/SSL overhead at the client and load at
the server. \spd additionally introduces the concept of \emph{priority}, which
allows the client to load important objects (\eg CSS and JavaScript) earlier,
\emph{server push}, which allows the server to push objects before the client
requests them, and \emph{header compression}, which reduces redundant
information shared between objects.

Many preliminary studies investigate \spd performance, but they fall short
either because they focus on an handful of websites~\cite{spdy_white}, or
explore a limited number of parameters~\cite{white_ietf, guypo, spdy_tech}.
Although the results of these studies are mostly contradictory, they indicate
poor \spd performance on mobile networks. Motivated by this observation, Erman
et al.~\cite{erman_2013CONEXT} set up \spd and \h proxies within a 3G network,
and monitored performance while fetching Alexa's top 20 websites They find that
\spd performs poorly in mobile networks due to many wasted retransmissions.

To address difficulties in measuring \spd performance, Xiao et
al.~\cite{wang_2014NSDI} devise a measurement platform to explore \spd
performance at a microscopic level. They find that \spd tends to outperform \h,
in the absence of browser dependencies and computation, however the gains are
reduced when both browser dependencies and computation are considered (with the
caveat that server push can squeeze additional performance gains from \spd).

In our work we also build a measurement platform and conduct experiments on 3G,
but we focus on \hh. More importantly, we do not need proxies since we target
real servers with \hh support, leading to different conclusions
than~\cite{erman_2013CONEXT}.\\

\noindent{\bf \hh} builds on \spd, though there are a few differences between
them. For example, \hh uses HPACK~\cite{hpack} for header compression,
eliminating \spd vulnerability to the ``crime" attack~\cite{crime},  and some
framing changes make \hh more regular than \spd.  De Saxc\`{e} et al. quantified
\hh performance in~\cite{saxce_2015GI}. Using the top 20 Alexa websites, but
stripping important components like the presence of multiple domains, they
investigate \hh performance while controlling network delay, bandwidth, and
losses via a measurement platform built with an open-source client and server.
Their results confirm previous observations about \spd in~\cite{wang_2014NSDI}.
We also investigate \hh performance but we use 600 real, unmodified \hh servers
discovered by monitoring the top 1 million Alexa sites, allowing us to gain
insights on \hh performance in the wild without ignoring widely used \h
optimizations like domain sharding and inlining.\\

\noindent{\bf NPN and ALPN} The Next Protocol Negotiation (NPN) \cite{npn}, an
extension  to Transport Layer Security (TLS) proposed by Google, is currently
used to  negotiate SPDY as an application level protocol over port 443 and to
perform \spd version negotiation. However, it is not \spd specific: servers
supporting NPN report which application layer protocols they handle, \eg~\hh~or
\spd,  and the client then selects which protocol to use. This negotiation is
done during the TLS  handshake without additional round trips. The Application-
Layer Protocol Negotiation  (ALPN) \cite{alpn} extension to TLS is an IETF
proposed standard set to supersede NPN.

\section{Measurement Platform}
\label{sec:mes}

This section describes our measurement platform, along with a collection of
tools we deployed for data collection.

\subsection{Tools}
\label{sec:mes:tools}

\noindent{\bf prober} is a lightweight bash script that identifies which
protocols a website announces. To do so, prober attempts an NPN negotiation and
reports either the list of protocols announced by the server or failure. NPN
negotiation is performed using OpenSSL~\cite{openssl}.  Next, prober checks for
\hh support over cleartext by including an UPGRADE header in an \h request.
Finally, prober checks if the site supports \quic, a Google transport protocol
designed for \spdy~\cite{quic}.  \quic support is advertised with an Alternate-
Protocol header in any response; we look for this header in the response to our
upgrade request.

\noindent{\bf h2-lite} is a tiny \hh client that attempts to download the root
object of a website using \hh. H2-lite uses the \hh library~\cite{molnarg}
implemented in Node.js~\cite{nodejs}. H2-lite will follow any redirects to
obtain the root object and report any protocol errors encountered along the way.

\noindent{\bf h2-full} is a headless browser that supports \hh, \spdy, and \h.
H2-full uses the Zombie.js headless browser~\cite{zombie} in conjunction with
the \hh library described above, a \spdy library~\cite{spdy_node} implemented in
Node.js, and the built-in Node.js \h library. H2-full can parse a website's root
object, recursively fetch embedded objects, and execute JavaScript. However only
a single protocol per run can be used, a limitation further discussed in
Section~\ref{sec:mes:lim}. H2-full reports metrics like page load time, number
of objects downloaded, number of domains present, number of TCP connections, and
object sizes. We also collect all HTML for offline analysis.

\noindent{\bf chrome-loader} is a tool written in Python that loads pages using
Chrome. It extracts object size and timing information using chrome-har-
capturer~\cite{har_capturer}. Unfortunately, \spdy and \hh cannot be
enabled/disabled independently, and since Chrome does not programmatically
report which protocol it used, we know only that \hh \emph{or} \spdy was used,
but not which.

\noindent{\bf page-saver} is a bash script that automates a browser's ``Save
Page As" feature~\cite{page_saver}. We use page-saver to collect the ground
truth of a web-page's composition, and compare it, when possible, with h2-full
and chrome-loader. Though any browser can be coupled with page-saver, we use
Chrome for consistency with the chrome-loader.

\subsection{Description}
\label{sec:mes:plat}

Our measurement platform consists of a \emph{master} and several \emph{workers}.
The master issues crawl requests to the workers, which are deployed both in
\pl~and in machines in our labs (both U.S. and Spain). We use \pl~for simple
measurements at large-scale, and our lab machines for more complex measurements
where reliable machines guarantee higher accuracy in data collection. Lab
machines are also useful when physical access to a machine is needed,  e.g., to
install a 3G dongle. Using a distributed system like \pl ensures that our
measurements are not (overly) biased by the geographic distance between clients
and the HTTP servers they are accessing. The master constantly monitors \pl to
identify a pool of ``reliable'' machines (at least 500~MB of free memory, CPU
load under 30\%, and no network connectivity issues).

Our measurement campaign consists of three phases, which we describe here in
detail.

\noindent{\bf Phase I:} First, the master launches an instance of the
\texttt{prober} at each reliable \pl worker. Each worker is then assigned an
unique 100-website list to probe.  Once a worker has probed each website on its
list, it reports results to the master, which stores them in a local database,
and obtains a new list if one is available. This approach ensures load balancing
among workers; faster workers tend to complete more tasks. To deal with
stragglers (slow workers), the master re-assigns uncompleted tasks to new
workers after a timeout $T$ set as the average task completion time across
workers.  Phase I terminates when we have determined which application layer
protocols are announced by the Alexa top 1M.

\noindent{\bf Phase II:} When Phase I terminates, the master extracts the list
of websites announcing \hh support. Then, it launches an instance of
\texttt{h2-lite} to fetch the root object from each website. When Phase II
terminates, the local database now includes a list of sites that \emph{actually
serve} content via \hh.

Because the \hh library requires more up-to-date software than is available on
\pl, we run Phase II on 3 servers under our control---2 in Barcelona, Spain and
1 in Cleveland, USA.  On each, we limit the number of concurrent website fetches
to 10.  Critically, Phase I limits the number of websites that must be explored
in Phase II to a small enough number so that our 3 servers can run Phase II
daily.

\noindent{\bf Phase III:} Once per week, Phase III runs on the list of websites
that Phase II has confirmed to deliver content over \hh. Each website is fetched
3 times each with \hh, \h, and \spdy (only if the website declared support in
Phase I). Phase III is run entirely from a single server in Barcelona, Spain for
uniformity of measurement apparatus across all websites and the website fetches
are run sequentially to limit the impact of network load on the results.

\subsection{Limitations}
\label{sec:mes:lim}

The main limitation of our measurement platform is the lack of support for ALPN.
This is due to the fact that ALPN support was added to OpenSSL only in version
1.0.2 released on January 22nd, 2015. Also, running the most recent version of
OpenSSL on \pl is tedious, due to a diverse and old set of operating systems.

In order to quantify the impact of this limitation, we instrumented one of our
lab machines to crawl the Alexa top 1 million websites looking for ALPN support.
We repeat this procedure once a month between February and April 2015, with the
goal to verify a potential switch from NPN to ALPN. This operation can be done
from a single machine in a couple of days, without the need of the massive \pl
scale. As of April 2015, only 0.5\% of the Alexa top 1 million websites support
ALPN, and they all support NPN as well. We plan to keep monitoring ALPN and
adapt our  measurement platform in the future to make sure it can react quickly
to ALPN eventual adoption.

The Node.js \spdy and \hh libraries do not support falling back to another
protocol if \spdy or \hh are not negotiated, respectively. Further, the  \spdy
library does not give any indication of which connection has failed when \spdy
cannot be negotiated. Thus, we are unable to fully download webpages if there
are embedded objects that are not accessible using the same protocol as the root
object. In the evaluation, we discuss how frequently we run into this
limitation and what counter-measures we take.

\begin{figure*}[t] 
\centering 
	\subfigure[Time evolution of application protocol support (left=general, right=\hh)]{\psfig{figure=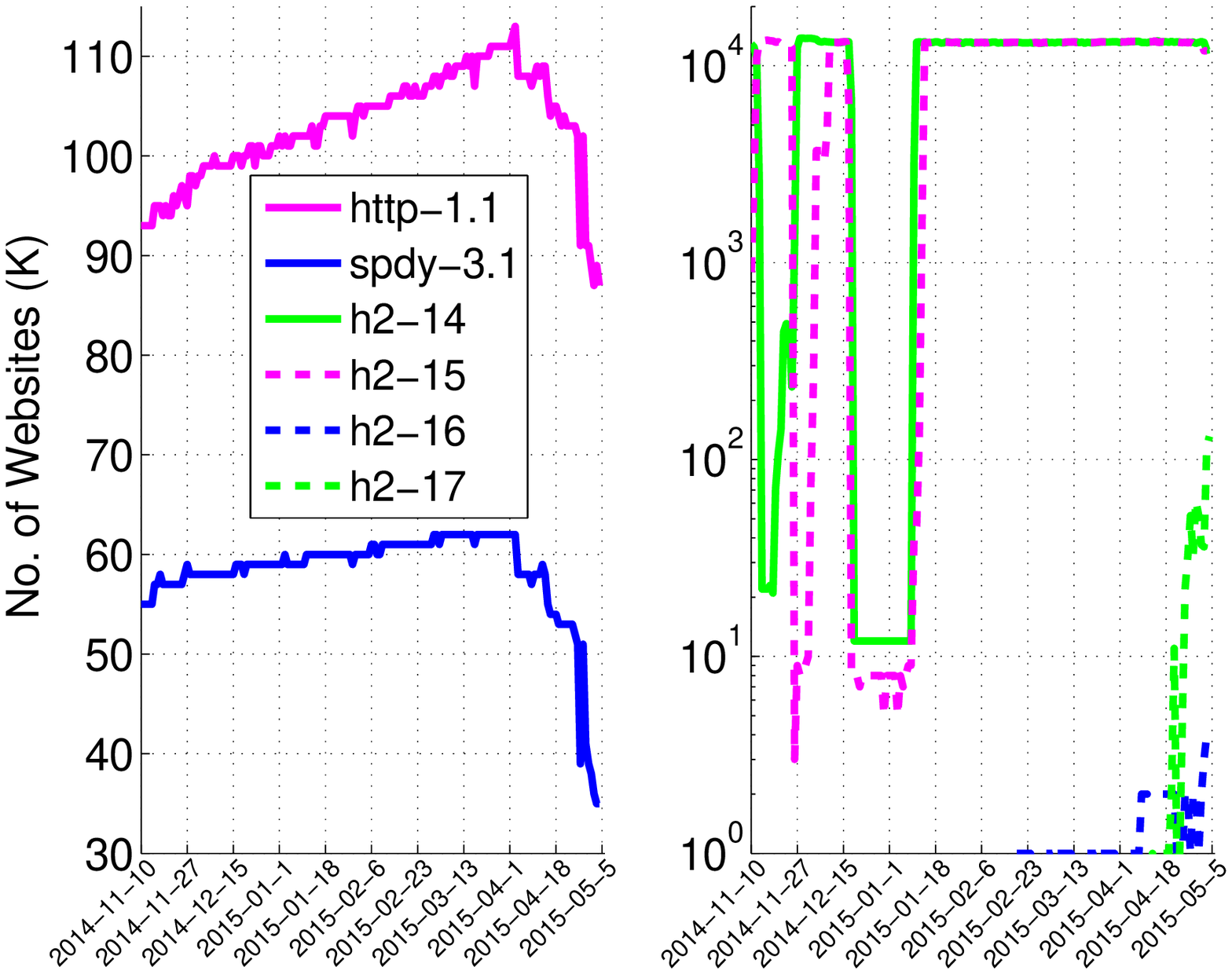,width=2.1in} 
	\label{fig:eval_time_1}}
	\subfigure[Time evolution of announced \hh~support vs actual support.]{\psfig{figure=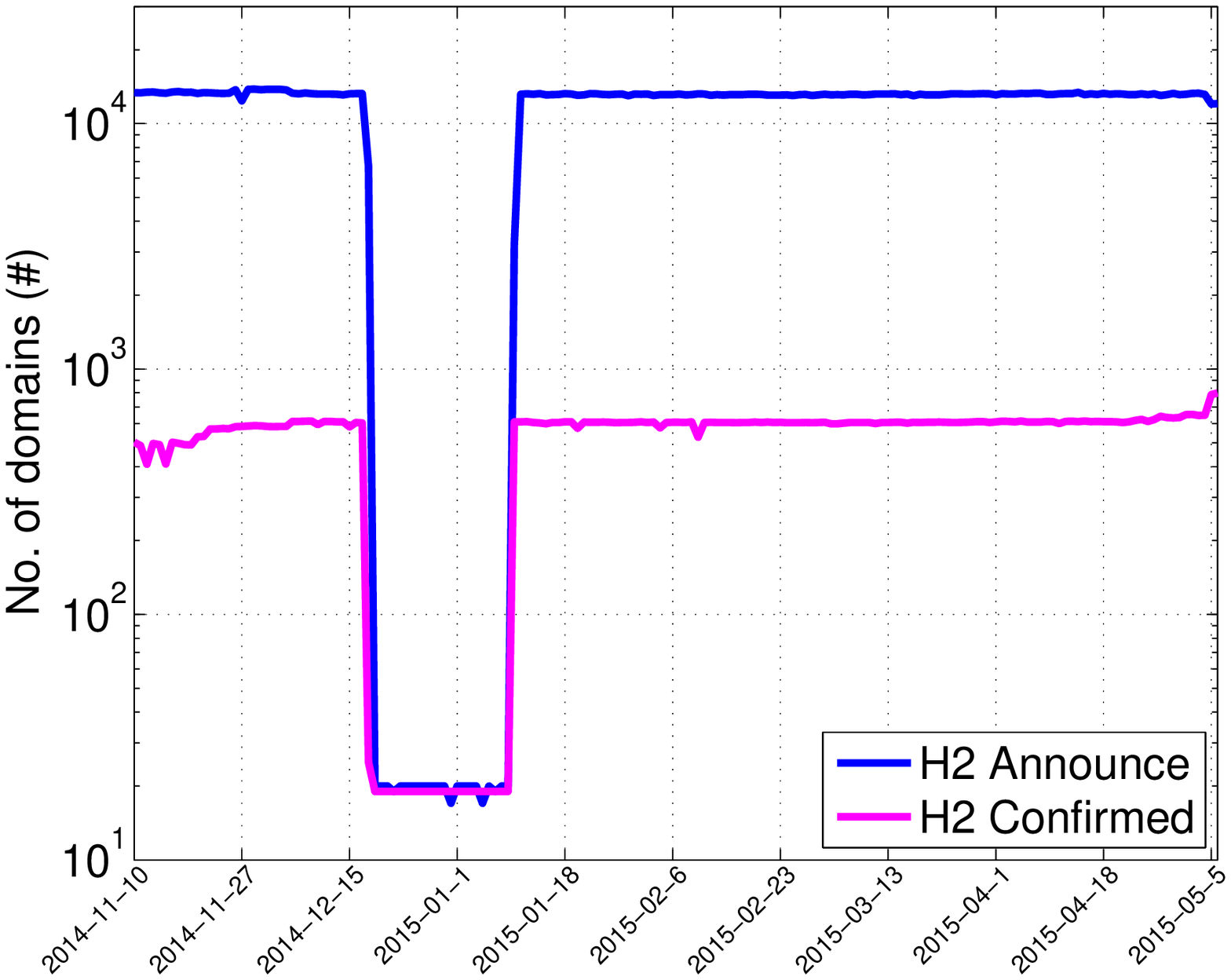, width=2.1in}
	\label{fig:eval_time_2}} 	
	\subfigure[Organization overview: \hh~vs \spd.]{\psfig{figure=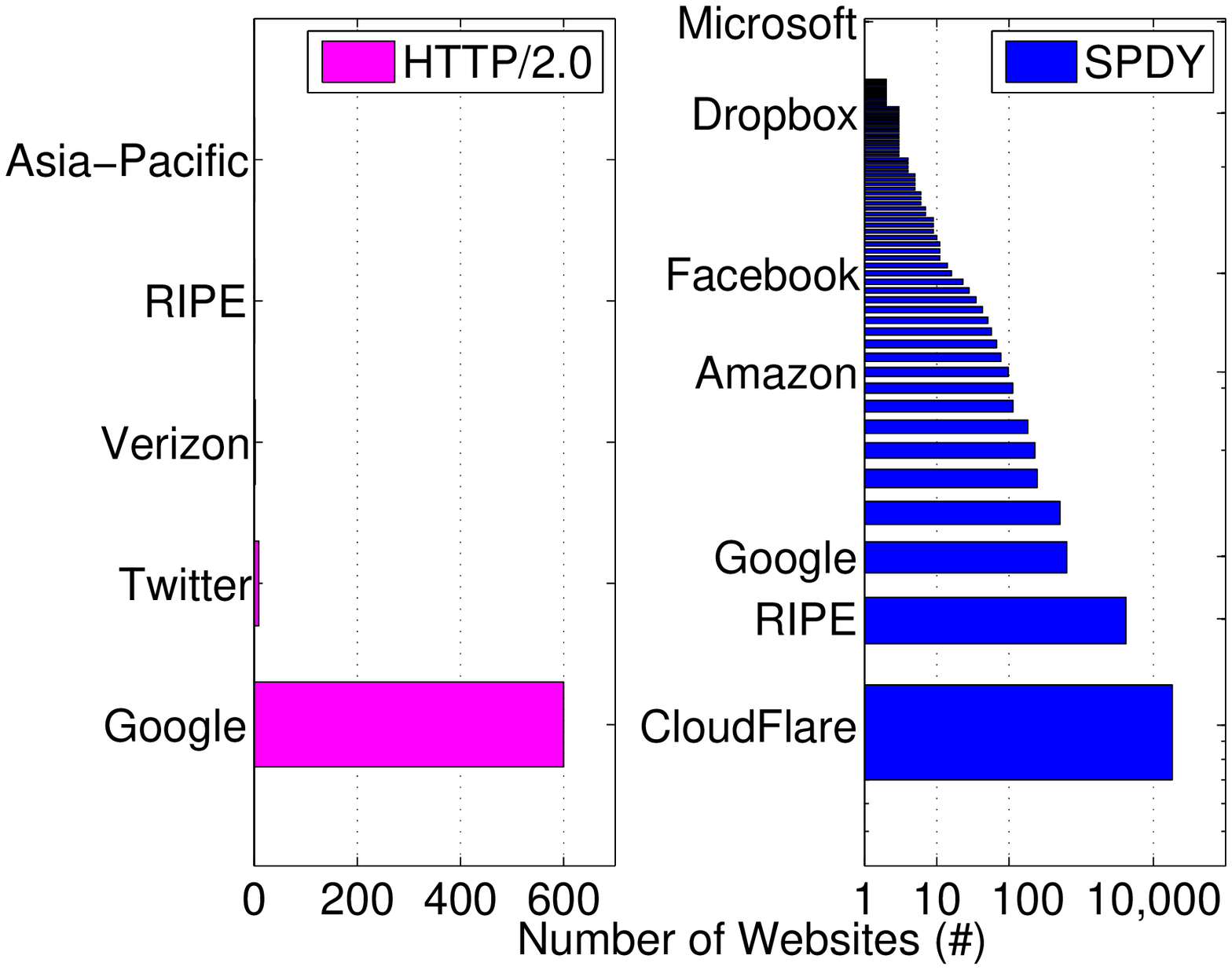,width=2.1in} 
	\label{fig:eval_time_3}}
\vspace{-0.15in}
\caption{\hh~Adoption.} 
\label{fig:eval_time} 
\vspace{-0.15in}
\end{figure*}

\section{Results}
\label{sec:results}

We leverage our measurement platform to collect traces  between November 10th,
2014 and May 6th, 2015. This section summarizes the analysis of the traces
collected. We invite the reader to access fresh data and analysis at \urldata.

\subsection{Adoption}
\label{sec:sec:long}

\begin{sloppypar}
Over six months---and about 200~million NPN-like negotiations---we have recorded
support for 44 application protocol versions, with \spd accounting for 34 of
them. \fig~\ref{fig:eval_time_1} shows the evolution over time of the fraction
of websites reporting to support either \texttt{http/1.1}, \texttt{spdy/3.1}, or
\texttt{h2} (cf.~Section~\ref{sec:background}), for which we distinguish between
versions \texttt{h2-14}, \texttt{h2-15},  \texttt{h2-16}, and \texttt{h2-17}.
\end{sloppypar}

\fig~\ref{fig:eval_time_1}(left) shows that, on November 10th 2014, about 92k
websites out of 1~million declared, via NPN, support for \h. This is an
underestimate of \h usage since \h does not require NPN. This measure really
reveals the number of websites supporting NPN. Accordingly the figure shows
that, over 5 months, NPN support grows by 20\%, from 92k to about 110k websites.
A similar trend is observable for \spd, where the number of supporting websites
goes from 55k to 62k. This is not surprising since NPN and SPDY are very much
coupled. However, in the last month we observed an opposite trend, with NPN
support dramatically dropping below where it was in November 2014. We verified
that this trend is not related to our measurement platform by testing the set of
websites that withdrew NPN support from two controlled machines. Note that the
introduction of ALPN does not explain this trend either (cf.~Section
\ref{sec:mes:lim}).

\fig~\ref{fig:eval_time_1}(right) focuses on \hh adoption only. Overall, the
plot shows about 13k websites declaring support, but negligible growth over 6
months.  Note that, to this date, not even a single website supports \hh over
cleartext. In November/December, most servers bounce back and forth between
announcing version h2-14 and h2-15. The most interesting event happens between
December 14th and 18th, when the number of servers supporting \hh~drops from 13k
to only 20 websites. This happens because Google, the major supporter of \hh
(cf.~\fig~\ref{fig:eval_time_3}) disabled \hh~support on their servers due to an
issue with SDCH compressed content in Chromium.\footnote{\small
https://lists.w3.org/Archives/Public/ietf-http-wg/2014OctDec/0960.html} It takes
a month for the problem to be fixed and \hh~support is re-enabled between
January 11th and the 15th. Interestingly, it takes Google four days to apply
this change to all its servers, probably reflecting a specific roll-out policy.
Between January and April 2015, \hh~support is quite stable. Version h2-16
appears first on February 19th, but it is not used by most servers. The final
\hh draft, h2-17, is released on February 11th, but support starts showing only
around April 16th, reaching 131 websites as of May 6, 2015.

We now focus on the 13k websites announcing \hh support and check whether they
actually serve content with it. Figure~\ref{fig:eval_time_2} shows the same
trend as in \fig~\ref{fig:eval_time_1},  with a large gap due to Google
suspending \hh~support temporally. The figure also shows that less than 5\% of
the websites  actually serve content through \hh; the remaining 95\% return code
301 (resource moved) and redirect \hh requests to \h. Only in the first month,
between November 10th and December 17th, is an actual growth of \hh support
noticeable, with 120 new websites starting to serve \hh content. None of these
websites has withdrawn support for \hh.

\fig~\ref{fig:eval_time_3} summarizes which organizations are behind \\
\hh~support. For a comparison, we also report the same information for \spd; out
of the 50-60k websites announcing \spd~support, about 23k actually serve some
content through it, if requested. Also, \quic is available at each of these
sites. Note that we do not report the evolution over time of \spd~announce and
support (dual of \fig~\ref{fig:eval_time_2}) because we started collecting this
information in the end of April 2015, and thus no interesting pattern is visible
yet. Overall, \fig~\ref{fig:eval_time_3} shows that only two major organizations
support \hh, Google and Twitter, with Google controlling about 99\% of the
websites. In comparison, the \spd~ecosystem is more complex, comprising 131
different organizations including CloudFlare and Facebook.

According to Alexa, 50\% of the websites supporting \hh today are in the top
100k websites. Also, three of the top 10 websites are present: \url{google.com},
\url{twitter.com}, and \url{youtube.com}. Thus, \emph{only a handful of sites
support \hh today, but those that do tend to be very popular}.

\begin{figure*}[t]
\centering
\subfigure[Number of Objects]{\psfig{figure=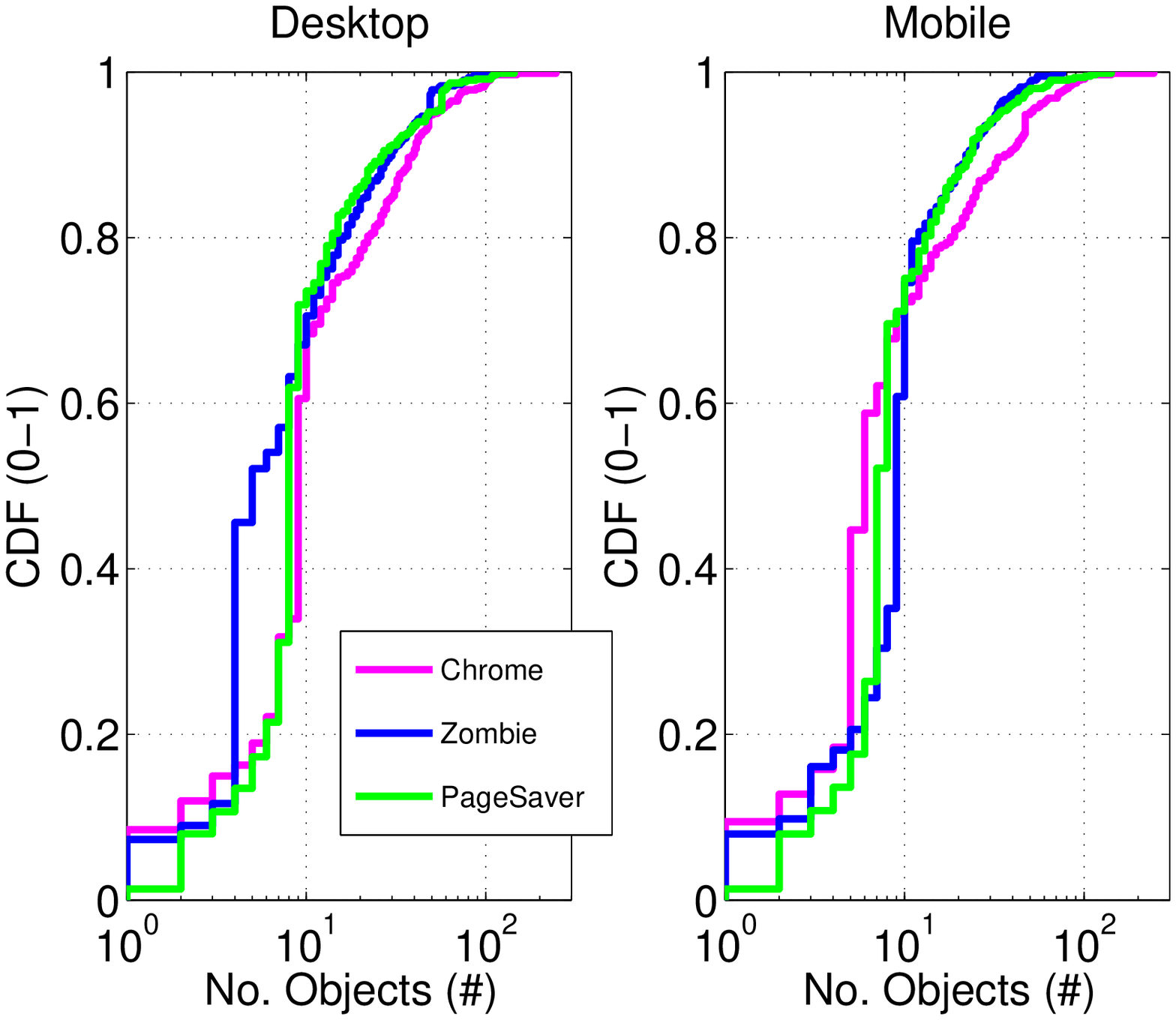, width=0.33\textwidth} \label{fig:macro_obj}}
\subfigure[Page composition]{\psfig{figure=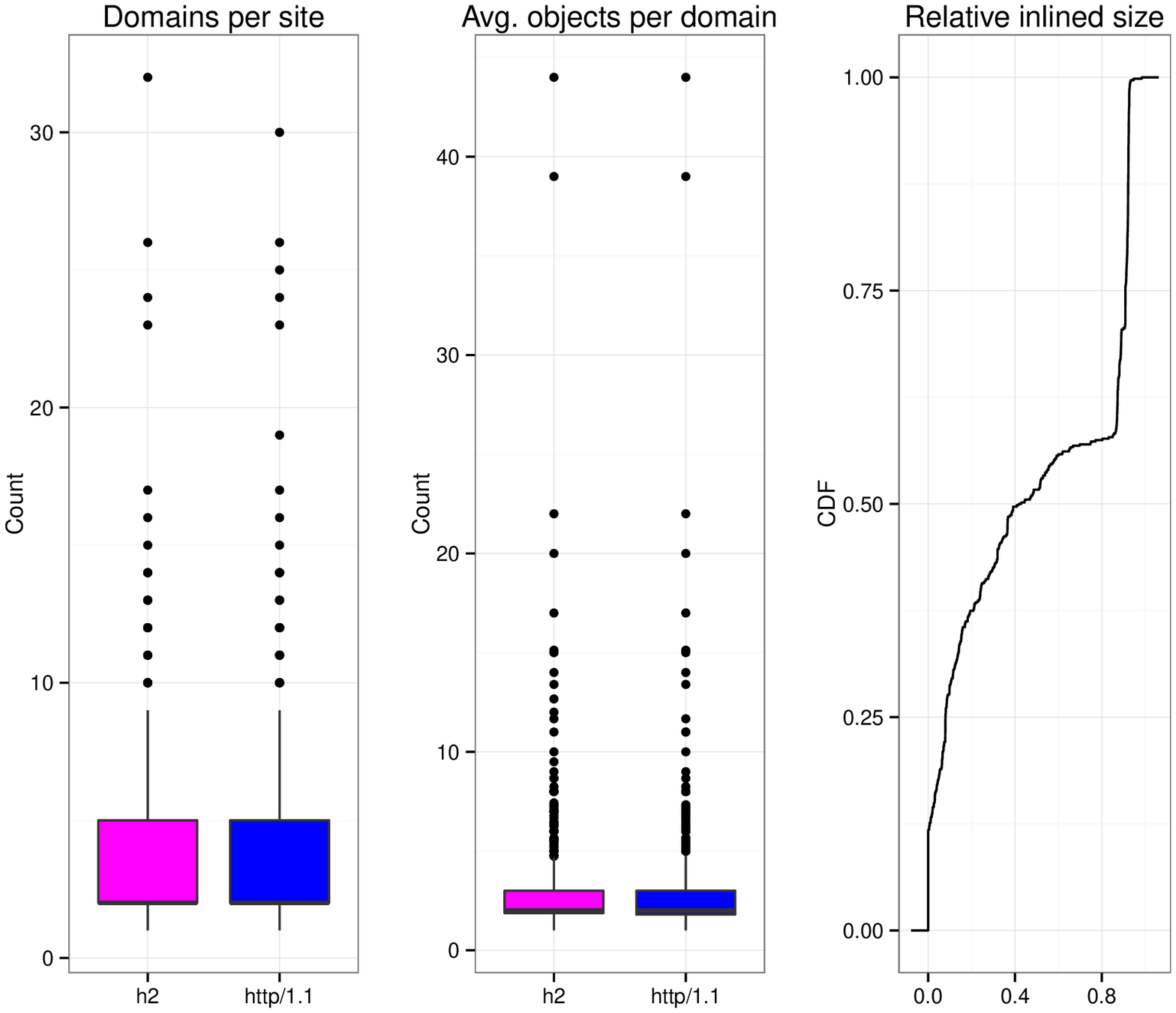, width=2.1in} \label{fig:microanalysis-boxplot}}
\subfigure[Number of Connections]{\psfig{figure=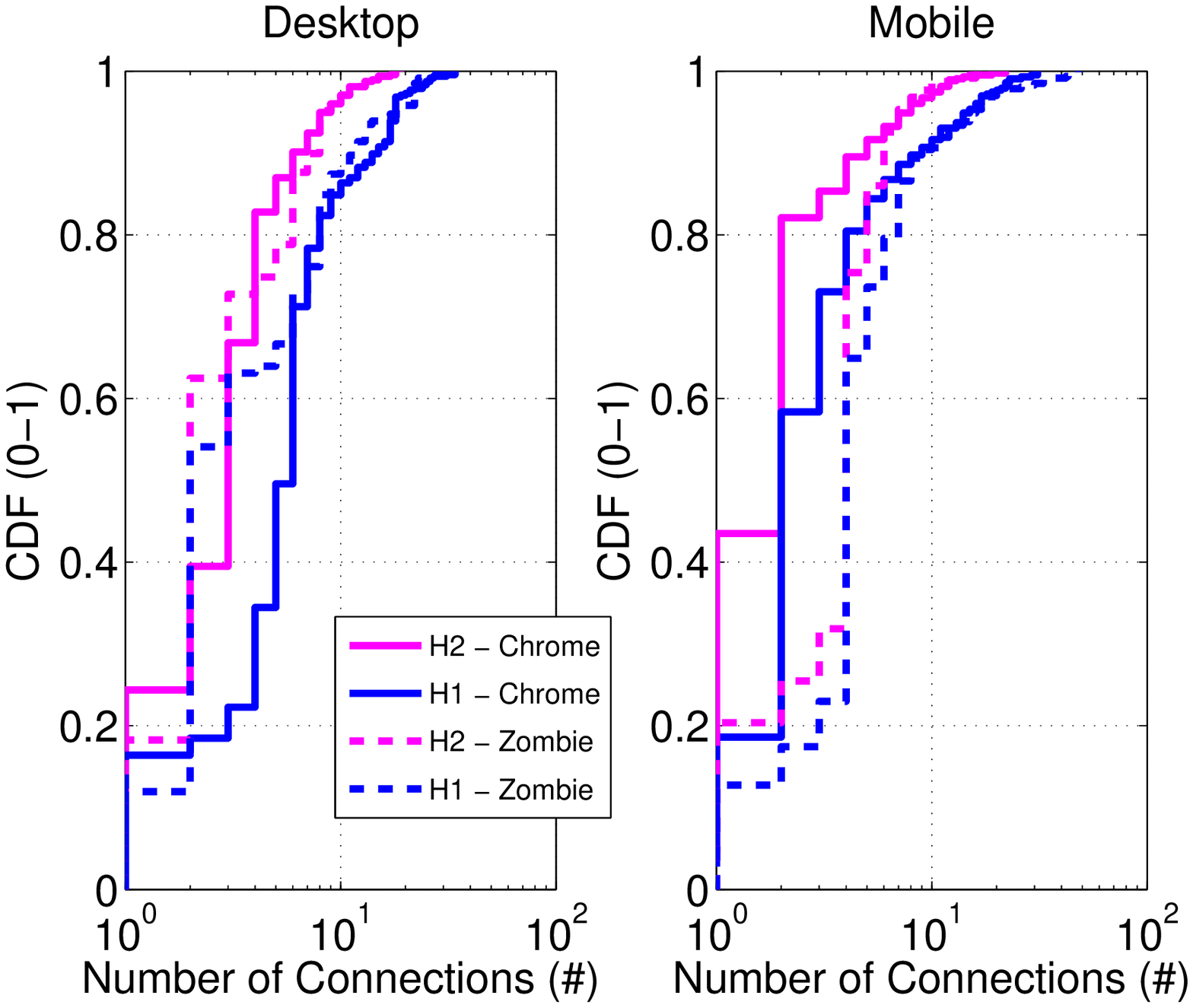, width=0.33\textwidth} \label{fig:macro_conn}}
\vspace{-0.15in}
\caption{Content analysis and delivery.}
\label{fig:macro}
\vspace{-0.15in}
\end{figure*}

\subsection{Content Analysis and Delivery}
\label{sec:sec:macro}

Next we investigate the composition of \spdy and \hh enabled sites (e.g., How
many objects do they embed? Do they exploit sharding and inlining?) and how
content is served (e.g., Do they use the same application protocol for all
objects? How many connections do they require?). For this analysis we focus on
April 30, 2015.

\noindent \textbf{Number of objects per website} \fig~\ref{fig:macro_obj} shows
the Cumulative Distribution Function (CDF) of the number of objects per webpage
when accessing them from a PC (desktop user agent) and a 3G dongle (mobile user
agent). Notice how different tools report very similar results in both
scenarios. The only macroscopic difference is related to Zombie under-estimating
the number of objects compared to both Chrome and page-saver for 40\% of the
websites.  These sites have a very similar composition; through manual
inspection, we find that Zombie fails to dynamically load objects created
through JavaScript or CSS embedded objects (e.g., background images).  In
particular, 70\% of sites where Zombie misses at least 4 objects are the
different country-specific Google homepages (e.g., \emph{google.es},
\emph{google.jp}). Such phenomenon is less visible in the mobile version where
this behavior is indeed less frequent.

Contrarily, Chrome seems to over-estimate the number of objects for about 20\%
of the sites while both Zombie and page-saver see similar figures. This is
curious since page-saver also uses Chrome.  Most of the sites are
\emph{*.appspot.com} and contain media files (e.g., .ogg, .woff2, or .glsl)
which Chrome does not export under the ``Save Page As'' function. Note that many
of these files are loaded by JavaScript which explains why Zombie misses them as
well. The takeaway is that \emph{web performance analysis is hard and using a
collection of tools helps gather meaningful results.}

Overall, most websites are composed of a small number of objects, \eg~80\% of
these sites only have about 20 objects, which reflects to an overall page of
less than 1~MB.  This is smaller than what we observe on the Web today, where,
on average, a page holds about 100 objects with an overall size of
2~MB~\cite{archive}.  This indicates that \emph{\hh~today is mostly serving
simple content, fewer in objects and size compared to the rest of the Web}.

\noindent\textbf{Page composition} Websites today try to optimize delivery using
a combination of \emph{sharding}, \emph{inlining}, \emph{spriting} and
\emph{concatenation}. Unfortunately, we see no straightforward way to measure
the latter two, so in this work we focus on the first two. Fig.~\ref{fig:microanalysis-boxplot} shows the results of this analysis.

Focusing first on sharding, the left boxplots in Fig.~\ref{fig:microanalysis-boxplot}(left) report the distribution of the number of unique domains per site.
Two observations hold. First, although the distributions for \h and \hh are
similar, we found several instances where a few objects were not loaded via \h
but were loaded by \hh. We conjecture this has to do with geo-location: when
accessing these sites via \h from Spain (but not the US) a few fewer objects are
returned. Second, we see that the number of domains per page has a median of 2,
a mean of 3.71, and a top 10th-percentile of 7 for both \h and \hh. Some sites
will open upwards of 30 connections \emph{solely because of a large number of
domains to contact}.

Next, in Fig.~\ref{fig:microanalysis-boxplot}(middle) we plot the distribution
of the average number of objects per domain per site. We observe that most
domains are used for relatively few objects (median = 2, mean = 2.96, top 10th-
percentile of 5.64, and a max of 44). This is further evidence of sharding-esque
behavior. \emph{Whether intentional or not, there are many sites that exhibit
the behavior we would expect from sharding}. This means many current websites
still require multiple TCP connections, even if they switch to \hh.

Finally, Fig.~\ref{fig:microanalysis-boxplot}(right) plots the size of inlined
CSS/JS to the total size of the main HTML for each site. We find that inlined
CSS/JS makes up over 80\% of the HTML content for over 40\% of the sites under
consideration. Although not shown in the plot, we further observed that CSS and
JavaScript inlining is prevalent: about 70\% of sites inline over 90\% of their
CSS and JavaScript objects. With \h, inlining can help ensure the page loads
faster by not requiring an extra TCP connection; however this is no longer an
issue in \hh, so \emph{these sites are (in theory) suffering a performance hit
since inlined content cannot be cached separately by the client}.

\noindent\textbf{Application protocol consistency} Even if a website supports
\spdy or \hh, embedded objects might be subject to different application
protocol negotiation. Indeed, we find that while 75\% of the sites serve 100\%
of their content with \hh, only 25\% serve the same amount with \spdy (and of
course 100\% of sites serve 100\% of content with \h). This result can be either
an artifact of the subset of sites we are investigating, chosen because of their
\hh rather than \spdy support, or of the \spdy library used.

\begin{sloppypar} 
\noindent\textbf{Number of connections} We analyze the number
of TCP connections used to load websites entirely delivered through \h or \hh.
\fig~\ref{fig:macro_conn} shows the CDF of the number of connections required by
Zombie and Chrome, per protocol. We only differentiate between \hh and \h since
we have no finer grained information for Chrome. No results are shown for page-
saver since it does not report connection counts. In the desktop version, \hh
with Zombie and \hh with Chrome have a similar trend with 78-81\% of web-pages
requiring 4 or fewer connections. Differences  in the number of objects per page
between Zombie and Chrome (cf.~\fig~\ref{fig:macro_obj})  cause Chrome to use
more connections. However, we also discovered an edge case in the \hh library
which allows multiple connections to the same domain upon multiple concurrent
requests for objects within the domain, explaining why Zombie rarely has more
connections  than Chrome. For \h, Chrome uses more connections for 55\% of web-
pages, again because  of differences in the number of objects per page. The
effect is more pronounced in \h  because Chrome allows up to 6 concurrent \h
connections per domain\cite{chrome_connections}. 
\end{sloppypar}

In the mobile version, we observe a dramatic difference in behavior: Chrome
requires fewer connections than Zombie across all protocols. First, Zombie
requires more  connections in the mobile version than in the desktop because
more objects are fetched in the mobile version (cf.~Fig.~\ref{fig:macro_obj}).
Second, we suspect that Chrome is under-reporting connections because for 40\%
of web-pages, Chrome reported opening fewer connections than the number of
domains which is, of course, impossible\footnote{Chrome likely reuses
connections from a previous download or preemptively opens connections before
they are needed.}. Note that this under-reporting occurs in the desktop version
as well and therefore  our measurements of connections used by Chrome are a
lower bound. If we now focus on differences between protocols measured by the
same tool, we notice that  \emph{\hh~reduces the number of TCP connections
compared to \h~by 10-15\%}.

\begin{figure}[t!]
\centering
\includegraphics[scale=0.3]{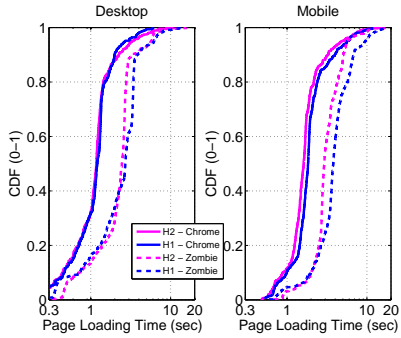}
\vspace{-0.15in}
\caption{CDF of page load time.}
\label{fig:plt_1}
\vspace{-0.15in}
\end{figure}

\noindent \textbf{Stability over time} Given the extensiveness of the analysis
(6 months), we further crosschecked the variation over time of the effects
described above. Specifically, we compute the cosine similarity for each metric
discussed both between successive weeks and between different protocols. Apart
from the number of connections, we find the cosine similarity always being
between 0.9 and 0.98, i.e., \emph{no significant content modification, at a
macroscopic level, as websites start adopting \hh}.

\subsection{Performance}
\label{sec:sec:perf}

We here investigate \hh speed in term of \emph{page load time (PLT)}.  There is
no standardized definition of PLT, so its meaning changes slightly between our
tools. In Zombie, PLT indicates the time elapsed between sending the initial
object request for a page and the ``visited'' event that signals Zombie has
finished the page. Note that not all of a page's objects must be downloaded
before a website  is considered ``finished.'' Similarly, PLT may be measured
from Chrome via the JavaScript ``onLoad'' event which indicates that the page's
resources have been loaded. Again, the ``onLoad'' event may occur before all
objects within the page are downloaded, since it does not apply to objects
loaded by scripts.

Because we measured no major content modification over 6 months
(cf.~Section~\ref{sec:sec:macro}), we focus on a single data point (April 30,
2015), differentiating between ``desktop'' (fiber + desktop user-agent) and
``mobile" (3G + mobile user-agent). Desktop and mobile data are collected from
our lab in Spain, using both Zombie and Chrome. Because only few of the sites we
are monitoring fully support \spdy, we focus the performance analysis on \hh and
\h only. Also, we consider only the subset of sites that can serve 100\% of
their content through \hh (477 desktop and 471 mobile).

Figure \ref{fig:plt_1}(left) shows the CDF of the average PLT (three downloads
per website) for the desktop scenario. Chrome reports a negligible difference
between PLT values obtained with \h and \hh (apart from the 10\% of websites
with PLT > 2 seconds) where \h saves on average $\sim$200~ms. If we focus on
Zombie results, we notice \hh being faster (between 100 and 500 ms) for half of
the sites whose PLT values fall between the 40th and 90th percentile. Note also
that Zombie shows much higher PLT values compared to Chrome, probably due to
Zombie being  far less optimized than Chrome.

\begin{figure}[t!]
\centering
\includegraphics[scale=0.3]{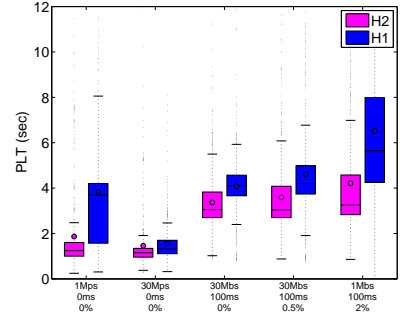}
\vspace{-0.15in}
\caption{PLT under controlled network configurations.}
\label{fig:plt_2}
\vspace{-0.15in}
\end{figure}

Next we consider PLT for mobile sites. Figure \ref{fig:plt_1}(right)  shows that
\hh consistently outperforms \h (according to both Chrome and Zombie). This
result is in contrast with~\cite{erman_2013CONEXT}, which suggests that \spdy,
which  is the basis of \hh, pays a large penalty in mobile. We do not observe
this behavior because our experiments do not require an \hh proxy, since we use
several real-world, deployed servers. Accordingly, we do not force a single TCP
connection as in~\cite{erman_2013CONEXT}, but open as many connections as needed
(cf.~\fig~ \ref{fig:macro_conn}), which is a side effect of spreading content
across multiple domains (cf.~\fig~\ref{fig:microanalysis-boxplot}). It follows
that \emph{domain sharding makes \hh---and likely \spdy---more resilient to
losses and delay variation typical of mobile networks.}

To further understand the previous results, we repeat the desktop scenario above
---Zombie only---while controlling network bandwidth, delay, and loss. \fig
\ref{fig:plt_2} shows the PLT distributions for both \h and \hh, for several
combinations of network parameters. We start by limiting the available bandwidth
to 1~Mbps; compared to the scenario with no bandwidth limitation (30~Mbps as
measured locally), we notice \h suffering a higher penalty in PLT compared to
\hh. A similar trend is noticeable as we create more challenging network
conditions, such as additional network delay and packet losses  (up to 100~ms
and 0.5\%).  Finally,  in the last scenario we limit the available bandwidth to
only 1~Mbps, and increase the packet losses to 2\%,  emulating a plausible
mobile scenario. The results confirm what was previously observed in
\fig~\ref{fig:plt_1}, with \hh greatly improving PLT compared to \h.

\section{Conclusion}
\label{sec:conclusion}

This work presents a measurement platform to monitor both adoption and
performance of \hh, the recently standardized update of \h. On a daily basis,
our platform checks for which protocols the top 1~million Alexa websites
announce support. Next, it checks which sites \emph{actually} support \hh.
Finally, once a week we benchmark \hh performance using the websites previously
discovered. Results are available at \urldata and are updated daily. In this
paper, we report our initial findings from the data collected between November
2014 and May 2015. We find 13,000 websites already announcing \hh support, out
of which 600 websites (mostly hosted by Google and Twitter) deliver content over
\hh.  A more in-depth analysis also unveils that classic \h hacks are still
present with \hh, and in challenging environments like 3G mobile, might result
in performance benefits.

\newpage
\bibliographystyle{acm}
\bibliography{BIBLIO}
\end{document}